\documentclass[twocolumn,secnumarabic,amssymb, bibnotes, aps, prl,reprint]{revtex4-1}
\usepackage{graphicx}

\begin{document}

\title{Damping of Confined Modes in a Ferromagnetic Thin Insulating Film: Angular Momentum Transfer Across a Nanoscale Field-defined Interface}%

\author{Rohan Adur}

\author{Chunhui Du}

\author{Hailong Wang}
\author{Sergei A. Manuilov}

\author{Vidya P. Bhallamudi}

\author{Chi Zhang}

\author{Denis V. Pelekhov}

\author{Fengyuan Yang}
\email{fyyang@physics.osu.edu}
\author{P. Chris Hammel}
\email{hammel@physics.osu.edu}
\affiliation{Department of Physics, The Ohio State University, Columbus OH 43210, USA}

\date{\today}

\begin{abstract}
We observe a dependence of the damping of a confined mode of precessing ferromagnetic magnetization on the size of the mode.  The micron-scale mode is created within an extended, unpatterned YIG film by means of the intense local dipolar field of a micromagnetic tip. We find that damping of the confined mode scales like the surface-to-volume ratio of the mode, indicating an interfacial damping effect (similar to spin pumping) due to the transfer of angular momentum from the confined mode to the spin sink of ferromagnetic material in the surrounding film. Though unexpected for insulating systems, the measured intralayer spin-mixing conductance $g_{\uparrow \downarrow} = 5.3 \times 10^{19} {\rm m}^{-2}$ demonstrates efficient intralayer angular momentum transfer.
\end{abstract}

\maketitle

Spin pumping driven by ferromagnetic resonance (FMR) is a powerful and well-established technique for generating pure spin currents in magnetic multilayers \cite{Tserkovnyak:prl02Damping,Brataas.prb02SpinBattery,Tserkovnyak:prl06VoltageGeneration,Heinrich:prlDynamicExchangeCoupling}. Understanding the mechanism that couples precessing magnetization to spin transport is an important step in utilizing this phenomenon. In addition, probing the effect of spin pumping on the damping of individual nanostructures is vital for the development of practical spintronic devices, such as spin-torque oscillators \cite{tsoi2000generation,kiselev2003microwave}. Conventional FMR studies at these sub-micron lengthscales become difficult due to the confounding effects arising from interfaces in multilayer materials and from sensitivity limitations in detecting lateral transport in single component systems at these length scales. Recent studies have shown that individual nanoscale elements exhibit size-dependent effects, such as nonlocal damping from edge modes \cite{Silva:nonlocal} and wavevector-dependent damping in perpendicular standing spin wave modes \cite{Bailey:k-dependent}. These experiments have revealed the effect of damping due to intralayer spin pumping, which is the transfer of angular momentum in systems with spatially-inhomogeneous dynamic magnetization.

A primary challenge in these measurements is distinguishing intralayer spin pumping from other mechanisms that cause variations in linewidth from sample to sample, such as surface and edge damage \cite{shaw:nanomagnets,noh:nanomagnets}. In this paper we measure size-dependent angular momentum transport across a clean interface without growth-defined defects or lithography-induced edge damage. This is achieved non-invasively in a single sample by confining the magnetization precession to a mode within an area defined by the controllable dipolar field from a nearby micron-sized magnetic particle \cite{h:lee.nature2010}. This enables a unique investigation of changes in relaxation due to angular momentum transfer across the field-defined interface between precessing magnetization within a mode to the spin sink provided by the surrounding quiescent material.

\begin{figure}
\includegraphics[width=\columnwidth]{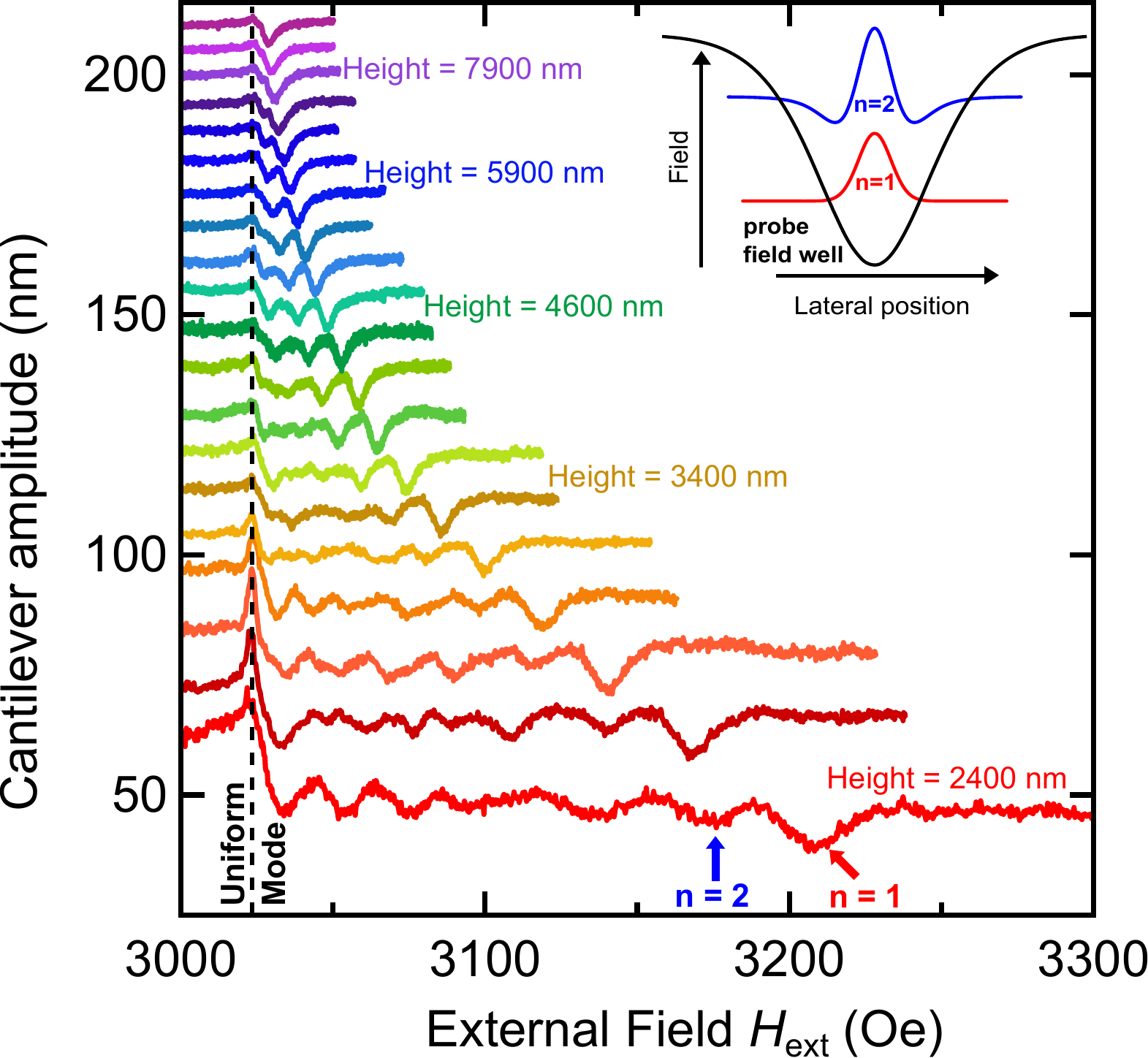}
\caption{Localized mode FMRFM spectra for thin film YIG at several probe-sample separations. The dashed line indicates the position of the uniform mode peak that does not shift with probe-sample separation. As probe-sample separation is reduced the localized modes shift to higher field relative to the uniform mode peak. Inset: transverse magnetization of the first two spin wave modes confined by the magnetic field well of the probe magnet. The energy of the confined modes is dictated by the depth of the field well.}
\label{Spectra_vs_Height}
\end{figure}

We investigate the size dependence of  interfacial damping using the technique of localized mode ferromagnetic resonance force microscopy (FMRFM) \cite{h:lee.nature2010}. By adjusting the magnitude of the dipolar field from the probe we can control the confinement radius. Localized modes have previously been observed in permalloy when the probe field is out-of-plane \cite{h:lee.nature2010}, in-plane \cite{McMichael:localizedmodes} and at intermediate angles \cite{h:nazaretski:PRBLocalFMR.2009}. The azimuthal symmetry of the out-of-plane geometry permits simple numerical analysis based on cylindrically symmetric Bessel function modes with a well-defined localization radius \cite{h:lee.nature2010}, similar to those seen in perpendicularly magnetized dots \cite{kakazei:aplDotArrays}. In addition, this geometry eliminates the effect of eigenmode splitting, which can cause additional broadening \cite{eason:nanomagnetsplitting}.

We demonstrate the control of confinement radius by the observation of discrete modes in an FMRFM experiment in the out-of-plane geometry in an unpatterned epitaxial yttrium iron garnet (YIG) film of thickness 25 nm grown by off-axis sputtering \cite{Wang:YIGspinpumping} on a (111)-oriented ${\rm Gd_3Ga_5O_{12}}$ substrate. The probe field is provided by a high coercivity ${\rm Sm_1Co_5}$ particle that is milled to 1.75 ${\rm \mu}$m after being mounted on an uncoated, diamond atomic force microscope cantilever. The magnetic moment and coercivity of the particle are measured by cantilever magnetometry to be ${3.9 \times 10^{-9}}$ emu and 10 kOe respectively. When the applied field is anti-parallel to the tip moment, the tip creates a confining field well in the sample that localizes discrete magnetization precession modes immediately beneath it \cite{h:lee.nature2010,Kalinikos:well}, analogous to the discrete modes in a quantum well \cite{fmr:Schlomann}. The microwave frequency magnetic field that excites the precession is provided by placing the sample near a short in a microstrip transmission line. A force-detected ferromagnetic resonance spectrum is obtained by modulating the amplitude of the microwaves at the cantilever frequency ($\approx$ 18 kHz) and measuring the change in cantilever amplitude as a function of swept external magnetic field. Measurements were made over a range of microwave frequencies:  2-6.5 GHz.

Figure \ref{Spectra_vs_Height} shows the evolution of the FMRFM spectra as a function of tip-sample separation obtained at a particular microwave frequency of 4 GHz. At large probe-sample separation we observe a peak at the expected resonance field for the uniform mode in the out-of-plane geometry. As expected, several discrete peaks emerge and shift toward higher applied field as the probe-sample separation decreases, thus increasing the (negative) probe field at the sample, while the uniform mode stays at constant resonance field. The resonance frequency $\omega$ of a confined mode for wavevectors $kt\ll 1$ is given by \cite{h:lee.nature2010}
\begin{equation}
\frac{\omega}{\gamma}=H_{\rm ext}-4 \pi M_s + \langle H_{p} \rangle+\pi M_s k t+ 4 \pi M_s  a_{ex}k^2
\label{resonance}
\end{equation}
where $\gamma = 2 \pi \times 2.8$ MHz/Oe is the gyromagnetic ratio, $H_{\rm ext}$ is the external applied magnetic field, $4 \pi M_s = 1608 $ Oe is the saturation magnetization, $a_{ex} = 3.6 \times 10^{-12} $ ${\rm cm^{2}}$ is the exchange constant of the material, $k$ is the wavevector of the mode, $t$ is the thickness of the film and $\langle H_{p} \rangle$ is the spatial average of the dipole field from the probe magnet weighted by the mode $m$
\begin{equation}
\langle H_{p} \rangle= \frac{\int_S H_p(r) \, m^2(r) \, {\mathrm d}^2 r}{\int_S m^2(r) \, {\mathrm d}^2 r}
\label{probefield}
\end{equation}

\begin{figure}
\includegraphics[width=\columnwidth]{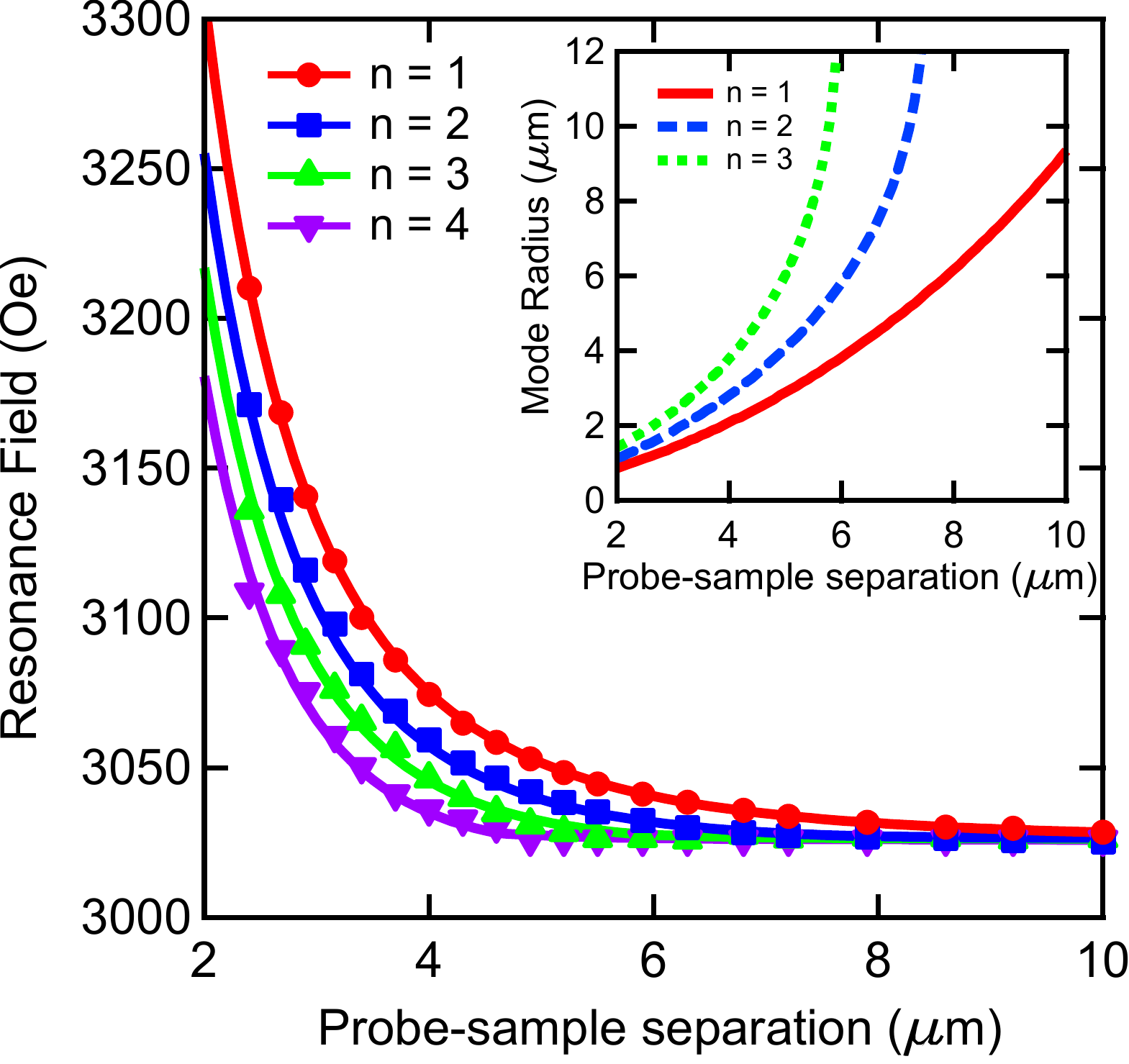}
\caption{Resonance fields of the first four localized modes as a function of probe-sample separation at 4 GHz. Filled markers indicate experimental peaks and solid lines indicate expected resonance field obtained numerically. Inset: radius of the first three localized modes obtained from the numerical minimization procedure described in the text.}
\label{Field_vs_Height}
\end{figure}

The film is sufficiently thin relative to the size of the probe particle that the dipole field is constant across the thickness of the film, and so the integration is performed over the sample surface $S$. Both the averaged probe field $\langle H_{p} \rangle$ and the wavevector $k$ are functions of the mode shape and mode radius $R$, so the frequency is obtained by numerical minimization with variation of radius \cite{h:lee.nature2010}. Due to cylindrical symmetry the magnetization profile of the mode can be described by a zeroth order Bessel function $m=J_0(kr)$, with boundary conditions that define discrete wavevectors $k_n = \chi_n/R$ where $\chi_n$ are the zeros of the Bessel function $ J_0(\chi_n) = 0$. The minimization of frequency at fixed field is equivalent to the maximization of field at fixed frequency. Hence the deeper field well shifts the modes to higher field when the microwave frequency is fixed, as seen in Fig.~\ref{Spectra_vs_Height}. This modeling procedure provides both the resonance field and the radius of the mode, and these are given in Fig.~\ref{Field_vs_Height}. We see that the resonance fields of the experimental peaks are well described by the model, confirming the accuracy of the calculated mode radius.

\begin{figure}
\includegraphics[width=\columnwidth]{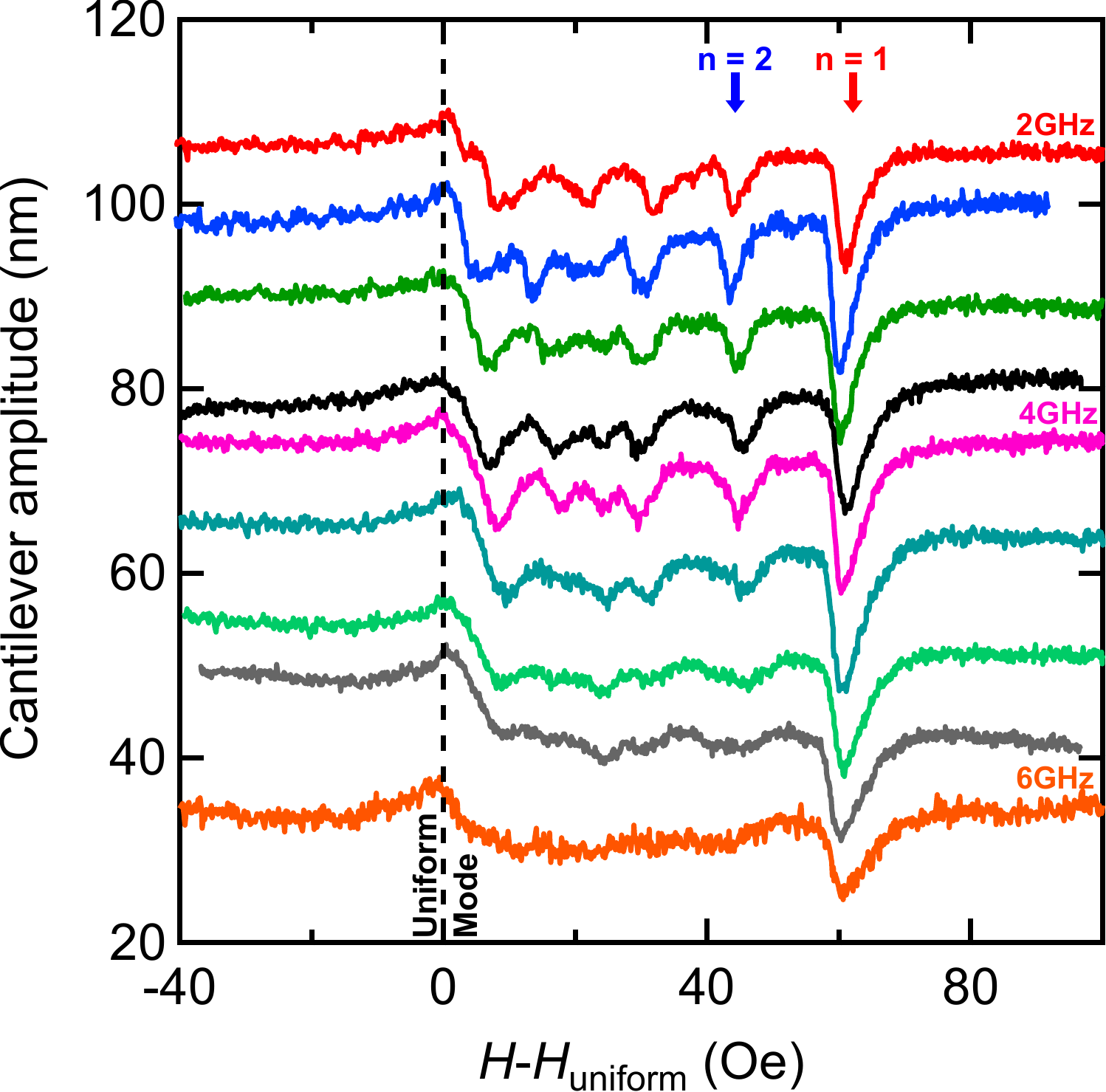}
\caption{FMRFM spectra at multiple microwave frequencies at a fixed probe-sample separation of 3700 nm, equivalent to a mode radius $R=$ 1860 nm. Spectra are offset for clarity and the external field $H$ is plotted relative to the uniform mode resonance field $ H_{\rm uniform}=\frac{\omega}{\gamma}+4\pi M_{\rm s}$.}
\label{Freq_Dependence_3700nm}
\end{figure}

To measure damping of a confined mode we obtain FMRFM spectra for a fixed mode radius $R$ at multiple frequencies, one example of which can be seen in Fig.~\ref{Freq_Dependence_3700nm}. The field shift of the localized modes, relative to the uniform mode $H_{\rm uniform}=\frac{\omega}{\gamma}+4\pi M_{\rm s}$ is constant for a fixed wavevector $k=k_n = \chi_n/R$, independent of frequency $\omega$, as predicted by Equation (\ref{resonance}).

By fitting a Lorentzian lineshape to the n = 1 and n = 2 peaks we obtain the full-width at half-maximum linewidth of the localized modes and plot this as a function of microwave frequency to separate intrinsic and extrinsic linewidth broadening mechanisms \cite{klein:FMRrelaxYig.prb2003}. Following from the Landau-Lifshitz-Gilbert equation, the linewidth $\Delta H$ is given by
\begin{equation}
\Delta H = \Delta H_0+\frac{2 \alpha \omega}{\gamma}
\label{linewidth}
\end{equation}
where the slope and intercept of the frequency-dependent linewidth measure, respectively, the Gilbert damping parameter $\alpha$ and inhomogeneous broadening $\Delta H_0$ due to spatial variation of magnetic properties. We measure this frequency dependence at several probe-sample separations corresponding to several mode radii $R$.

\begin{figure}
\includegraphics[width=\columnwidth]{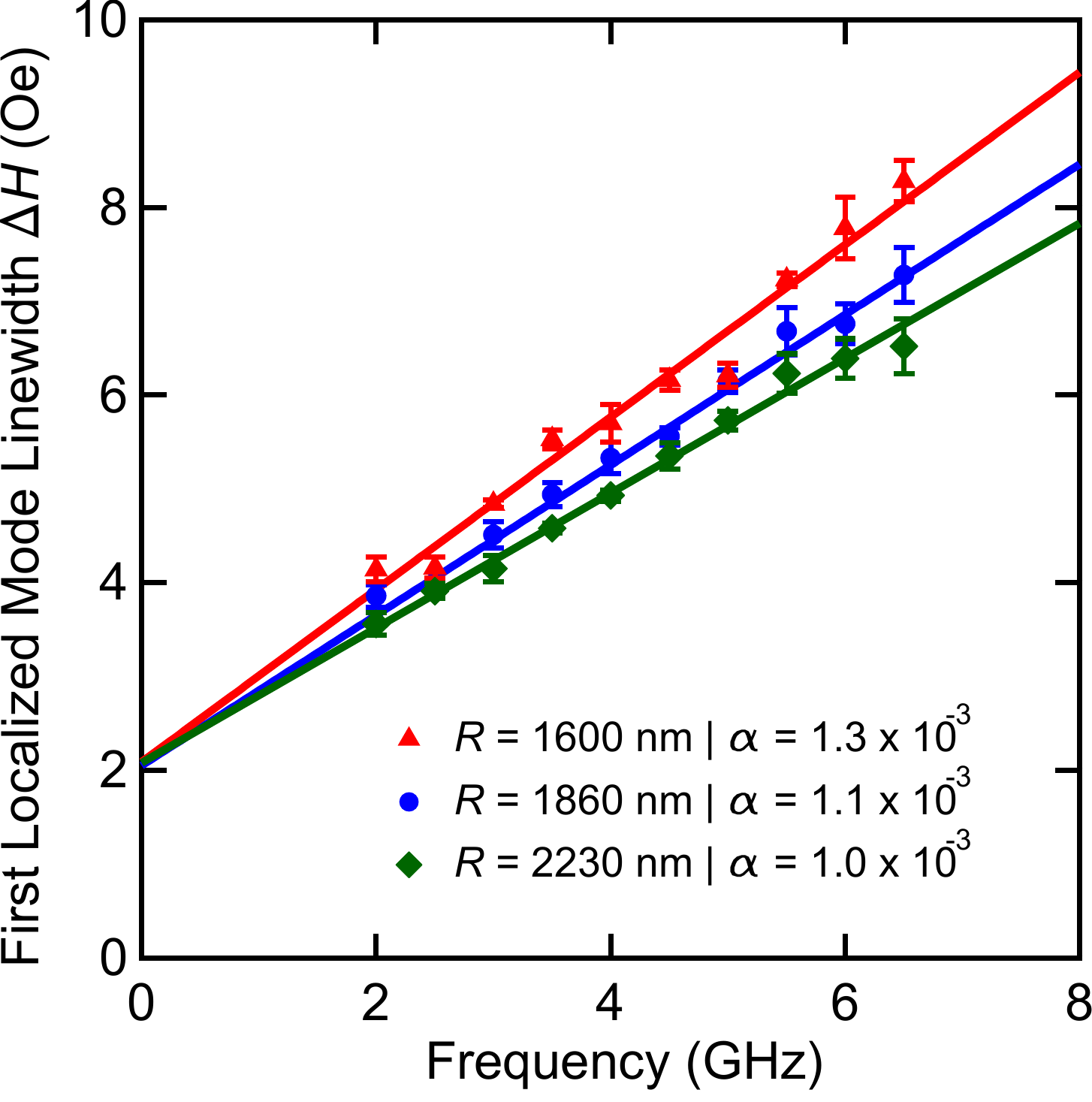}
\caption{Linewidths of first localized mode for mode radii $R=$ 1600 nm (red triangles), $R=$ 1860 nm (blue squares) and $R =$ 2230 nm (green diamonds). Filled markers are experimental linewidths and solid lines are linear fits to the data. Gilbert damping parameters $\alpha$ are determined for each mode radius from the slope of the linear fit.}
\label{freqlinewidth}
\end{figure}

\begin{figure}
\includegraphics[width=\columnwidth]{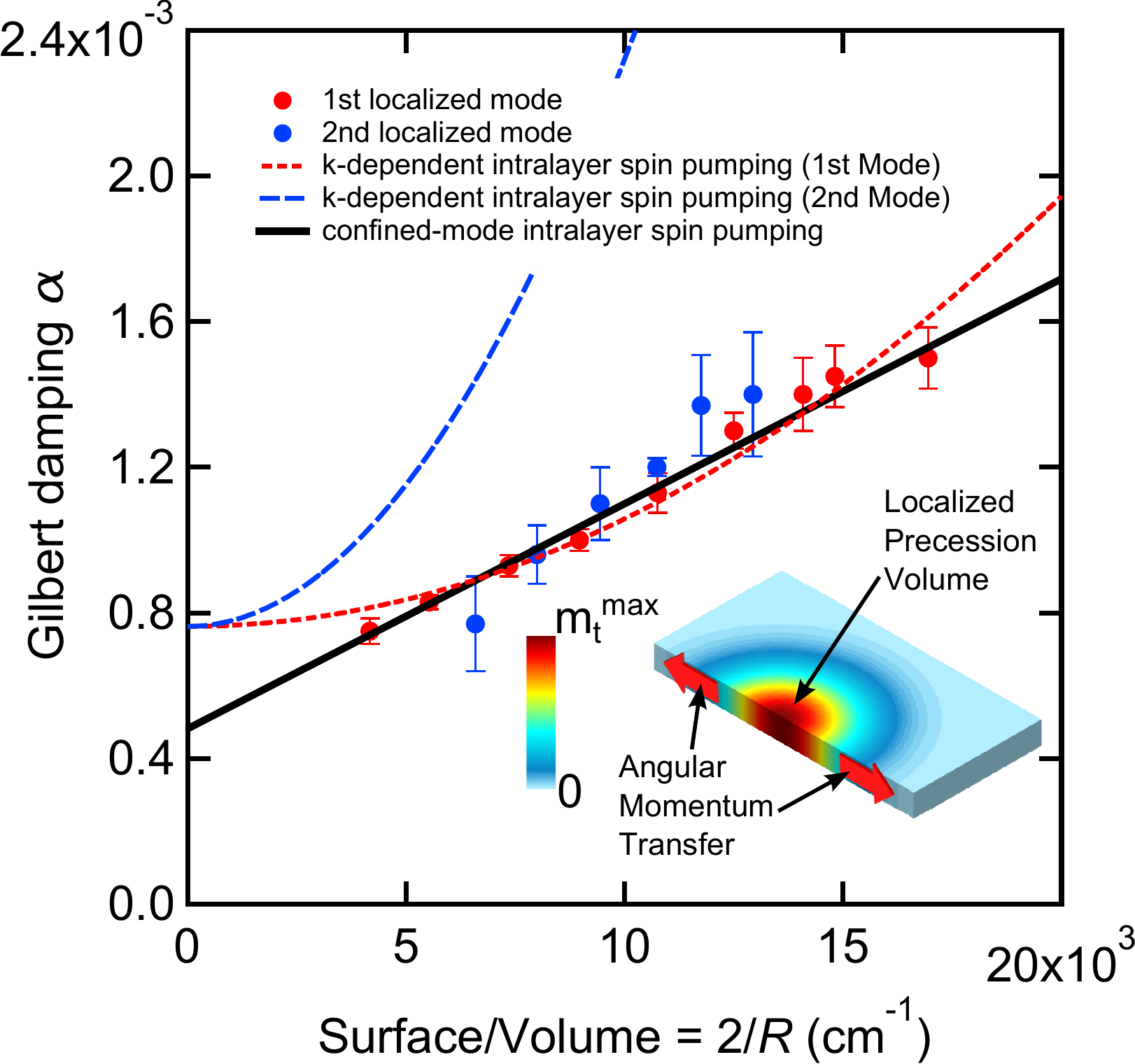}
\caption{Comparison of the measured size-dependent Gilbert damping parameter $\alpha$ of the first two localized modes with theory. The solid black line is a linear fit to confined-mode intralayer spin pumping that scales with the surface-volume ratio of the mode as described by Eq.~(\ref{spinpumpdisc}). The dashed red line is a fit to the first localized mode linewidth using wavevector-dependent intralayer spin pumping theory \cite{TserkovnyakVignale} that scales as $k^2$ \cite{Silva:nonlocal}. The dashed blue line is the prediction of wavevector-dependent intralayer spin pumping for the second localized mode.  Inset: Cross-section showing intralayer angular momentum transfer from the volume of the confined mode to surrounding material through the surface of the mode. Color scale denotes magnitude of the transverse, precessing magnetization $m_t$.}
\label{alpha_vs_radius}
\end{figure}

The key result of our study is the observation of enhanced damping that is unambiguously dependent on the radius of the mode, as seen from the change in slope of the first localized mode linewidth with mode radius as seen in Fig.~\ref{freqlinewidth}. The Gilbert damping parameter $\alpha$, for both the first and second localized modes, shows a surprising linear behavior when plotted against $R^{-1}$, the reciprocal of the mode radius, as seen in Fig.~\ref{alpha_vs_radius}. An enhanced damping is reminiscent of spin pumping observed when a ferromagnetic layer is placed in contact with a normal metal layer \cite{Tserkovnyak:prl02Damping}. In this bilayer geometry the damping enhancement $\alpha_{\rm sp}$ scales inversely with thickness $t$ of the FM film, which is equal to the ratio of the area of the ferromagnet/metal interface to the volume of the ferromagnet, and is given by \cite{heinrich:YIGAuFespinpumping}
\begin{equation}
\alpha_{\rm sp} = \frac{\gamma \hbar g_{\uparrow \downarrow}}{4 \pi M_s} \frac{1}{t}
\label{spinpumpfilm}
\end{equation}
where $\hbar$ is the reduced Planck constant and $g_{\uparrow \downarrow}$ is the spin-mixing conductance parameter that describes the efficiency of spin pumping. By analogy to this interfacial damping due to spin pumping we suggest the possibility of an interfacial damping mechanism for confined modes that scales with the surface-to-volume ratio of the mode, where the volume of the on-resonant disk-like mode is $\pi R^2 t$ and relaxation to the surrounding material, which is off resonance, occurs through the curved surface $2 \pi R t$ around the edge of the disc. Hence, the enhanced damping of a confined mode with radius $R$ is
\begin{equation}
\alpha_{\rm sp} = \frac{\gamma \hbar g_{\uparrow \downarrow}}{4 \pi M_s} \frac{2}{R}
\label{spinpumpdisc}
\end{equation}
From Equation (\ref{spinpumpdisc}) and the linear fit (solid black line) to the enhanced damping versus mode the reciprocal of the mode radius, as shown in Fig.~\ref{alpha_vs_radius}, we obtain $g_{\uparrow \downarrow} = (5.3  \pm 0.2) \times 10^{19} {\rm m}^{-2}$ for this system.

It is interesting and somewhat remarkable that we observe angular momentum transport in this insulating system and that its efficiency, characterized by $g_{\uparrow \downarrow}$, is larger than the spin-mixing conductance measured in YIG-metal bilayers \cite{bauer:backflow,YIGmetalscaling,Wang:YIGspinpumping}. We suggest that $g_{\uparrow \downarrow}$ measured in this study is an \textit{intralayer} spin-mixing conductance that describes a generalization of spin pumping as the transport of energy and angular momentum from an on-resonance spin source to an off-resonance spin sink, even in the absence of both a material interface \cite{Silva:nonlocal} and conduction electrons \cite{hahn:insulatorspinpumping}. We describe this effect as YIG-YIG intralayer spin pumping: the energy and angular momentum from the precessing confined mode can be absorbed by the surrounding ferromagnetic material of the unpatterned film, as depicted in the inset of Fig.~\ref{alpha_vs_radius}. The relatively large value of $g_{\uparrow \downarrow} = (5.3  \pm 0.2) \times 10^{19} {\rm m}^{-2}$ we obtain for YIG-YIG can be compared to $g_{\uparrow \downarrow} = (6.9 \pm 0.6) \times 10^{18} {\rm m}^{-2}$ previously measured for YIG-Pt \cite{YIGmetalscaling}. This enhancement may arise because the interface, rather than involving a material discontinuity, is defined by a magnetic field that occurs in a uniform, essentially defect-free film leading to a strong "interfacial" coupling characterized by the YIG-YIG exchange interaction itself. In addition, it might be unexpected for the confined mode to relax via the surrounding material where the lowest energy state, which is the uniform mode, is well above the energy of the confined mode inside the well. However, previous experiments by Heinrich et al.~\cite{Heinrich:prlDynamicExchangeCoupling,heinrich:YIGAuFespinpumping} have shown that ferromagnets do act as good spin sinks when the precession frequency of the spin current source is not at a resonance frequency of the spin sink ferromagnet.

We consider the possible role of transverse spin diffusion \cite{TserkovnyakVignale} used previously to describe enhanced damping due to the interaction between itinerant electrons and spatially-inhomogeneous dynamic magnetization \cite{Silva:nonlocal,Bailey:k-dependent}. We find that the prediction of this wavevector-dependent intralayer spin pumping theory does not agree with our experimental data. In particular, this enhanced damping due to intralayer spin pumping is predicted \cite{TserkovnyakVignale} to depend on wavevector k:
\begin{equation}
\alpha_{\rm sp}=\frac{\sigma_T \gamma}{M_s} k^2
\label{intralayer}
\end{equation}
where $\sigma_T$ is the transverse spin conductivity and the wavevector $k = \chi_n/R$ is given by the Bessel zeros $\chi_1$ = 2.405, $\chi_2$ = 5.520. The spin conductivity due to itinerant electrons is expected to be zero in YIG, but we nevertheless allow it to be a free parameter and fit to the first localized mode linewidth; this fit to the wavevector-dependent intralayer spin pumping theory is shown as the red dashed line in Fig.~\ref{alpha_vs_radius}. We find that the spin conductivity that describes this fit, $\sigma_T = 1.5 \times 10^{-22}$ kg m/s, is two orders of magnitude larger than that measured in a metallic ferromagnet \cite{Silva:nonlocal}. In addition, using the same spin conductivity to estimate the linewidth of the second localized mode (blue dashed line) results in a prediction that does not accurately describe the measured second mode linewidth (blue solid circles), while confined-mode intralayer spin pumping that scales as the surface-volume ratio of the mode (black solid line) described by Eq.~(\ref{spinpumpdisc}) accurately describes both sets of data. Hence our observations do not follow the wavevector-dependent intralayer spin pumping theory observed previously \cite{Silva:nonlocal,Bailey:k-dependent}, but manifests as a surface-volume intralayer relaxation specific to spatially-confined precession within an extended film, previously predicted for nanocontact spin-torque oscillators \cite{Slavin:nanocontact}.

Other mechanisms for linewidth broadening are ruled out by analysis of the phenomenology of our result. The dipolar field from the micromagnetic tip is a potential source of linewidth broadening as it is produces an inhomogneous field in the sample of several hundred gauss that would dominate inhomogeneous spectral broadening in a paramagnetic sample \cite{h:MagHandbook,suter:probesample}. Inhomogeneous broadening from the tip can be ruled out as the source of increased damping in this study for two reasons. First, any inhomogeneous broadening would be frequency independent, and hence would lead to a change in the intercept of the frequency-dependence of linewidth shown in Fig.~\ref{freqlinewidth}, while the change in slope alone is a clear indication of a Gilbert damping enhancement. Second, the ferromagnetic resonance excitations of a ferromagnet are eigenmodes \cite{h:lee.nature2010,h:MagHandbook}, in which the inhomogeneous field from the tip is cancelled by the dynamic field from the precession. This allows the effective field to be equal at every position inside the mode, and hence it can be described as an eigenmode with a single well-defined eigenfrequency. Other well-established mechanisms for size- or wavevector-dependent relaxation can also be eliminated due to their insufficient magnitude and differing phenomenology; 3-magnon confluence \cite{kasuya-lecraw,sparks} manifests as a linewidth broadening that is linear in $k$ but independent of frequency, while 4-magnon scattering \cite{Dyson} scales as $k^2$.

To conclude, we observe robust intralayer spin pumping within an insulating ferromagnet, which manifests as enhanced damping of micrometer-scale confined spin wave modes. This result has consequences for devices that induce spin precession in confined regions, such as spin-torque oscillators in the nanocontact geometry \cite{madami:spintorquewave,demidov:spinhalloscillator}. In addition, our study highlights the power of localized mode FMRFM for illuminating local spin dynamics and in particular for spectroscopic studies of the impact of mode relaxation across a controllable, field-defined interface.

The authors wish to thank Yaroslav Tserkovnyak for useful discussions. This work was primarily supported by the U.S. Department of Energy (DOE), Office of Science, Basic Energy Sciences (BES), under Award \# DE-FG02-03ER46054 (FMRFM measurement) and Award \# DE-SC0001304 (sample synthesis). This work was partially supported by the Center for Emergent Materials, an NSF-funded MRSEC under award \# DMR-0820414 (structural characterization). This work was supported in part by Lake Shore Cryotronics (magnetic characterization) and an allocation of computing time from the Ohio Supercomputer Center (micromagnetic simulations). We also acknowledge technical support and assistance provided by the NanoSystems Laboratory at the Ohio State University.

\end{document}